\documentclass[superscriptaddress,twocolumn,nofootinbib,floatfix]{revtex4-1}

\usepackage{graphicx}
\usepackage{epsfig}
\usepackage{amsmath,amssymb}
\usepackage{booktabs}
\usepackage{color}

\newcommand{\cbeta}{c_\beta}
\newcommand{\sbeta}{s_\beta}
\newcommand{\calpha}{c_\alpha}
\newcommand{\salpha}{s_\alpha}
\newcommand{\cW}{c_W}
\newcommand{\sW}{s_W}
\newcommand{\cZ}{c_Z}
\newcommand{\sZ}{s_Z}

\begin{document}

\title{Multiboson signals in the UN2HDM}

\author{J.~A.~Aguilar-Saavedra}
\affiliation{Instituto de F\'{\i}sica Te\'{o}rica UAM-CSIC, Campus de Cantoblanco, E-28049 Madrid, Spain}
\author{F. R. Joaquim}
\affiliation{Departamento de F\'{\i}sica and CFTP, Instituto Superior T\'ecnico, Universidade de Lisboa, Av. Rovisco Pais 1, 1049-001 Lisboa, Portugal}
\author{J. F. Seabra}
\affiliation{Instituto de F\'{\i}sica Te\'{o}rica UAM-CSIC, Campus de Cantoblanco, E-28049 Madrid, Spain}
\affiliation{Departamento de F\'{\i}sica and CFTP, Instituto Superior T\'ecnico, Universidade de Lisboa, Av. Rovisco Pais 1, 1049-001 Lisboa, Portugal}


\begin{abstract}
We address multiboson production from a heavy $Z'$ resonance in the context of the UN2HDM, a standard model extension with an additional $\text{U}(1)'$ symmetry and an enlarged scalar sector with an extra doublet and a singlet. After taking into account theoretical and experimental constraints on the model, it turns out this type of signals --- mostly uncovered by current searches --- could be sizeable. We focus on three benchmark scenarios, each of them predicting up to 4000 multiboson events with the LHC Run 2 collected luminosity. Anomaly-detection methods could uncover those signals, if present in data.
\end{abstract}

\maketitle

\section{Introduction}

Despite the belief that there must be physics beyond the Standard Model (SM), the way it may manifest at collider experiments is yet to be understood. This circumstance reinforces the need to put forward new physics signals that searches at the Large Hadron Collider (LHC) could have been missing, as well as models in which those signals are produced. An example of such uncovered signal is multiboson production arising from the cascade decay of a new resonance~\cite{Aguilar-Saavedra:2015rna,Aguilar-Saavedra:2015iew}. Such signal, namely a triboson resonance, was proposed as alternative interpretation of a $3.4\sigma$ bump near 2 TeV in an ATLAS search for hadronically-decaying diboson resonances with Run 1 data~\cite{ATLAS:2015xom}. Although searches for tri-$W$ resonances have been performed by the CMS Collaboration~\cite{CMS:2022lqh}, focusing on models with extra dimensions~\cite{Agashe:2018leo,Agashe:2017wss}, more general triboson as well as quadriboson resonance signals~\cite{Aguilar-Saavedra:2017iso} are not experimentally covered.

The persistence of small bumps near 2 TeV in diboson searches using Run 2 data, with a small local significance of $2\sigma$ both in the ATLAS~\cite{ATLAS:2016yqq} and CMS~\cite{CMS:2017fgc} experiments, motivated another uncovered signature, dubbed as `stealth boson'~\cite{Aguilar-Saavedra:2017zuc}. This is a boosted particle (generically denoted as $S$) with a cascade decay
\begin{equation}
S\rightarrow A_1 A_2\rightarrow q\overline{q}q\overline{q} \,, 
\label{eq:SB}
\end{equation}
where the intermediate particles $A_{1,2}$ can be SM weak bosons ($W$ and $Z$), the SM Higgs boson, or new scalars. When $S$ is produced with a high boost from the decay of a much heavier resonance $R$, the products of its hadronic decay are reconstructed as a single jet with four-pronged structure. Multiboson signals involving cascade decays such as in Eq.~(\ref{eq:SB}) can give rise to merged four-pronged jets, and the complex jet structure makes the resulting signals quite more elusive for current searches.

Minimal extensions of the SM allowing hadronic cascade decays of stealth bosons were proposed in~\cite{Aguilar-Saavedra:2019adu} and further explored in~\cite{Aguilar-Saavedra:2020wrj}. They were called minimal stealth boson models (MSBMs) and assume that the heavy resonance $R$ is a colour-singlet neutral gauge boson $Z'$. In order to break the corresponding ${\rm U(1)}'$ symmetry and simultaneously account for the cascade decay in Eq.~(\ref{eq:SB}), MSBMs include two complex scalar singlets, together with extra matter that is required to cancel ${\rm U(1)}'$ anomalies.

In this work we explore an alternative model that can give rise to these types of signatures. Our study is especially timely given the latest bump near 2 TeV reported by the CMS collaboration in the search for hadronically-decaying diboson resonances with the full Run 2 dataset, reaching a local significance of $3.6\sigma$~\cite{CMS:2022shx}.
The model we present in this work also introduces a heavy gauge boson $Z'$, but replaces one of the scalar singlets of MSBMs by a complex scalar doublet. We label the model as UN2HDM, as it corresponds to a next-to-two-Higgs doublet model (N2HDM)~\cite{Chen:2013jvg,Muhlleitner:2016mzt,Keus:2017ioh,vonBuddenbrock:2018xar,Ferreira:2019iqb,Biekotter:2019kde,Engeln:2020fld}  with an extra $\text{U}(1)'$ symmetry. In comparison with MSBMs, the presence of a larger scalar sector in the UN2HDM leads to the presence of charged scalar particles. A detailed overview of the UN2HDM and the interactions of the new fields is presented in Section~\ref{sec:2}.

Our main goal is to show that, in the context of the UN2HDM, multiboson signals can be produced with sizeable cross sections in regions of the parameter space allowed by theoretical and experimental constraints. We generically denote the signals as multibosons, including also the case when the bosons are very boosted and their hadronic decay products merge into a single jet. In the notation of Eq.~\eqref{eq:SB}, $S$ will be $Z'$, and $A_{1,2}$ two neutral or charged scalars, which subsequently undergo direct or cascade decays into boosted jets. After explaining in Section~\ref{sec:3} how we scan the parameter space to find points allowed by constraints, we study in Section~\ref{sec:4} three benchmark scenarios for multiboson signals. We discuss our results in Section~\ref{sec:5}.

\section{The UN2HDM}
\label{sec:2}
\begin{table*}[t!]
\centering
\setlength{\tabcolsep}{20pt}
\begin{tabular}{ccc}
 SM fermions & Vector-like leptons & Scalars \\ \midrule \addlinespace[0.1cm]
$q_L = \begin{pmatrix} u & d \end{pmatrix}_L\sim (1/6,Y'_q)$
& $\mathcal{E}_L\equiv\begin{pmatrix} N_1 & E_1 \end{pmatrix}_{L}\sim (-1/2,-9 Y'_q/2)$
& $\Phi_1\sim (1/2,9Y_q')$ \\\addlinespace[0.1cm]
$u_R \sim (2/3,Y'_q)$ 
& $\mathcal{E}_R\equiv\begin{pmatrix} N_1 & E_1 \end{pmatrix}_{R}\sim (-1/2,9 Y'_q/2)$
& $\Phi_2\sim (1/2,0)$ \\\addlinespace[0.1cm]
$d_R \sim (-1/3,Y'_q)$
& ${N_2}_{L}\sim (0,9 Y'_q/2)$
& $\chi\sim (0,9Y_q')$ \\\addlinespace[0.1cm]
$\ell_L = \begin{pmatrix} \nu_l & l \end{pmatrix}_L\sim (-1/2,0)$ 
& ${N_2}_{R}\sim (0,-9 Y'_q/2)$
&  \\\addlinespace[0.1cm]
$l_R \sim (-1,0)$ 
& ${E_2}_{L}\sim (-1,9 Y'_q/2)$
& \\\addlinespace[0.1cm]
& ${E_2}_{R}\sim (-1,-9 Y'_q/2)$
& 
\end{tabular}
\caption{SM and U(1)' hypercharges  $\psi\sim(Y,Y')$ of the fermions and scalars in the UN2HDM. Generation indices in SM fermions are omitted, as well as colour indices for quarks.}
\label{table:Hypercharges}
\end{table*}
The UN2HDM extends the SM gauge group with an extra U(1)$'$ symmetry, thus featuring a new colour-singlet neutral gauge boson $Z'$. Given the strong limits from $Z'$ boson searches in their leptonic decays, $Z'\rightarrow e^+e^-$ and $Z'\rightarrow \mu^+\mu^-$, we consider the new $Z'$ boson to be leptophobic, with the U(1)$'$ hypercharges of SM lepton doublets and singlets vanishing, i.e. $Y'_\ell=Y'_e=0$. As already mentioned, the UN2HDM scalar sector contains two doublets $\Phi_1$ and $\Phi_2$, and one complex singlet $\chi$. Cascade decays like the one in Eq.~(\ref{eq:SB}) cannot take place if both doublets have vanishing U(1)$'$ hypercharge. On the other hand, due to the requirements of U(1)$'$ gauge invariance of the Yukawa terms of SM leptons, and of a leptophobic $Z'$, one of the Higgs doublets must have zero hypercharge under U(1)$'$. We therefore label the scalar doublets so that $Y'_{\Phi_2} = 0$ and $Y'_{\Phi_1} \neq 0$. The U(1)$'$ gauge invariance of the quark Yukawa terms with $\Phi_2$ implies $Y'_q = Y'_u= Y'_d$, generally non-vanishing. Therefore, we have the same Yukawa lagrangian found in a Type I two-Higgs doublet model (2HDM), namely
\begin{equation}
-\mathcal{L}_Y = Y_u\overline{q}_L\tilde{\Phi}_2u_R + Y_d\overline{q}_L\Phi_2 d_R + Y_e\overline{\ell}_L\Phi_2 e_R + {\rm h.c.}\,,
\end{equation}
where $Y_X\,(X=u,d,e)$ are complex Yukawa matrices in generation space. The U(1)$'$ hypercharges of SM fermions are identical to those of MSBMs. Thus, all U(1)$'$ anomalies~\cite{Langacker:2008yv} cancel by adding the same extra matter as in MSBMs. Assuming the extra matter to be vector-like under the SM group, two simple solutions have been proposed in ref.~\cite{Aguilar-Saavedra:2019adu}: one with a set of vector-like quarks (Model 1), and another with a set of vector-like leptons (Model 2). In this work we will concentrate only on Model 2, for which the field content and hypercharge assignments are given in Table~\ref{table:Hypercharges}. The new vector-like leptons are the two SU(2) doublets $(N_1\,E_1)_{L,R}$ and the four singlets ${N_2}_{L,R}$ and ${E_2}_{L,R}$. $N_i$ and $E_i$ have electric charge 0 and -1, respectively. Their hypercharges are fixed by anomaly cancellation, and all take values of $ \pm 9 Y'_q / 2$. Gauge invariance of the Yukawa interactions of the new leptons with the scalar singlet (which give rise to their masses) implies
\begin{equation}
Y_\chi'=9Y_q' \,,
\end{equation}
leading to the following Yukawa interactions involving vector-like leptons:
\begin{equation}
    \begin{aligned}
    \mathcal{L}_Y^{\rm VLL} &= \overline{\mathcal{E}}_L(w_1^N\tilde{\Phi}_2{N_2}_R + w_1^E\Phi_2{E_2}_R) \\
    &+ (w_2^N\overline{N_2}_R\tilde{\Phi}_2^\dagger + w_2^E\overline{E_2}_R\Phi_2^\dagger)\mathcal{E}_R \\
    &+ y_1^N\overline{N_2}_L{N_2}_R\chi + y_1^E\overline{E_2}_L{E_2}_R\chi \\ 
    &+ y_2\overline{\mathcal{E}}_L\mathcal{E}_R\chi^\dagger +{\rm h.c.}\,,
    \end{aligned}
    \label{eq: YukawasVLL}
\end{equation}
with $w_i^{N,E}$ and $y_i^{N,E}$ ($i=1,2$) being complex Yukawa couplings.
We also assume a `dark lepton number' that forbids Majorana mass terms for $N_{2L}$ and $N_{2R}$~\cite{Caron:2018yzp}. Finally, as we will see later, a valid scalar mass spectrum can only be obtained if $Y_{\Phi_1}'=Y_\chi'$.

With these $\text{U}(1)'$ hypercharge assignments, the most general gauge-invariant scalar potential of the UN2HDM is
\begin{align}
& V = m_{11}^2\Phi_1^\dagger\Phi_1 + m_{22}^2\Phi_2^\dagger\Phi_2 + \frac{m_0^2}{2}\chi^\dagger\chi \notag \\ 
& \quad + \frac{\lambda_1}{2}(\Phi_1^\dagger\Phi_1)^2 + \frac{\lambda_2}{2}(\Phi_2^\dagger\Phi_2)^2 
+ \lambda_3(\Phi_1^\dagger\Phi_1)(\Phi_2^\dagger\Phi_2) \notag \\
& \quad + \lambda_4(\Phi_1^\dagger\Phi_2)(\Phi_2^\dagger\Phi_1) + \frac{\lambda_5}{2}(\chi^\dagger\chi)^2 + \frac{\lambda_6}{2}(\Phi_1^\dagger\Phi_1)(\chi^\dagger\chi) \notag \\ 
& \quad + \frac{\lambda_7}{2}(\Phi_2^\dagger\Phi_2)(\chi^\dagger\chi) + (\,\mu\chi\Phi_1^\dagger\Phi_2+ {\rm h.c.})\,,
\label{eq:V}
\end{align}
where $\mu$ can be a complex parameter, while the remaining parameters are real. We define the scalar doublets $\Phi_{1,2}$ and singlet $\chi$ as 
\begin{equation}
\begin{gathered}
    \Phi_k = \frac{1}{\sqrt{2}}\begin{pmatrix} \sqrt{2}\phi_k^+ \\ v_ke^{i\varphi_k} + \rho_k + i\eta_k \end{pmatrix}\,,\quad (k=1,2)\,, \\
    \chi = \frac{1}{\sqrt{2}}(ue^{i\varphi_3} + \rho_3 + i\eta_3)\,.
        \end{gathered}
\end{equation}
Without loss of generality, one can assume $\varphi_1=0$, such that the vacuum expectation values (VEVs) are
\begin{equation}
    \begin{gathered}
    \langle\Phi_1\rangle = \frac{1}{\sqrt{2}}\begin{pmatrix} 0 \\ v_1 \end{pmatrix}\,,\quad
    \langle\Phi_2\rangle = \frac{1}{\sqrt{2}}\begin{pmatrix} 0 \\ v_2e^{i\varphi_2} \end{pmatrix}\,\quad \\
    \langle\chi\rangle = \frac{1}{\sqrt{2}}ue^{i\varphi_3}.
    \end{gathered}
\end{equation}
Throughout this work, we will always assume nonzero VEVs $v_1$, $v_2$ and $u$. Note also that $\varphi_2$ and $\varphi_3$ can be rephased away through 
\begin{equation}
\Phi_2\rightarrow \Phi_2'=e^{-i\varphi_2}\Phi_2 \,,\quad
\chi\rightarrow \chi'=e^{-i\varphi_3}\chi\,,
\end{equation}
which leaves $V$ invariant, provided $\mu$ is replaced by
\begin{equation}
\mu\rightarrow \mu'=\mu e^{-i(\varphi_2+\varphi_3)}\,,
\end{equation}
when $V$ is expressed in terms of $\Phi_2'$ and $\chi'$. Since $\mu$ is an arbitrary complex parameter, from now on we assume $\varphi_2=\varphi_3=0$ without loss of generality, therefore having all VEVs real.

An useful feature of $V$ is that the number of parameters is equal to the number of physical quantities (masses and mixing angles) needed to define the Higgs sector. This means that all eleven parameters shown in Eq.~(\ref{eq:V}) can be written in terms of the three VEVs $v_1$, $v_2$ and $u$, as well as of the five scalar masses and three mixing angles that will be introduced later on.

For the cases we are interested in with $v_1,v_2,u\neq 0$, the four minimisation conditions of the scalar potential are
\begin{align}
& m_{11}^2 + \frac{v_1^2}{2}\lambda_1 + \frac{v_2^2}{2}(\lambda_3 + \lambda_4) + \frac{u^2}{4}\lambda_6 + \frac{u}{\sqrt{2}}\frac{v_2}{v_1}\Re(\mu)=0\,, \notag \\
& m_{22}^2 + \frac{v_2^2}{2}\lambda_2 +\frac{v_1^2}{2}(\lambda_3 + \lambda_4) + \frac{u^2}{4}\lambda_7 + \frac{u}{\sqrt{2}}\frac{v_1}{v_2}\Re(\mu)=0\,, \notag \\
& m_0^2 + u^2\lambda_5 + \frac{1}{2}v_1^2\lambda_6 + \frac{1}{2}v_2^2\lambda_7 + \sqrt{2}\frac{v_1v_2}{u}\Re(\mu)=0\,, \notag \\
& \Im(\mu)=0\,.
\label{eq:mincond}
\end{align}
The first three equations allow to write $m_{11}^2$, $m_{22}^2$ and $m_0^2$ in terms of the VEVs and the remaining parameters of the potential.
The last equation in (\ref{eq:mincond}) implies that all parameters in the potential are real and, thus, there are no mixed $\rho_i \eta_j$ mass terms. Then, the $6\times 6$ neutral scalar mass matrix can be written in block-diagonal form as
\begin{equation}
\mathcal{M}^n = \begin{pmatrix} M^\rho & 0 \\ 0 & M^\eta \end{pmatrix}\,,
\end{equation}
where $M^\rho$ and $M^\eta$ are $3\times 3$ real symmetric matrices defined in the $(\rho_1,\rho_2,\rho_3)$ and $(\eta_1,\eta_2,\eta_3)$ bases, respectively.
Using henceforth the definitions
\begin{equation}
v = \sqrt{v_1^2 + v_2^2}\quad,\quad \tan\beta = v_2/v_1\,,
\end{equation}
and the notation $s_\beta = \sin\beta$, $c_\beta = \cos\beta$, the independent elements of $M^\rho$ read
\begin{align}
& M^\rho_{11} = v^2\lambda_1\cbeta^2 - \frac{1}{\sqrt{2}}u\mu\tan\beta\,, \notag \\
& M^\rho_{12} = v^2(\lambda_3+\lambda_4)\sbeta\cbeta + \frac{1}{\sqrt{2}}u\mu\,, \notag \\
& M^\rho_{13} = \frac{v}{2}u\lambda_6\cbeta + \frac{1}{\sqrt{2}}v\mu\sbeta\,, \notag \\
& M^\rho_{22} = v^2\lambda_2\sbeta^2 - \frac{u\mu}{\sqrt{2}\tan\beta}\,, \notag \\
& M^\rho_{23} = \frac{v}{2}u\lambda_7\sbeta + \frac{1}{\sqrt{2}}v\mu\cbeta\,, \notag \\
& M^\rho_{33} = \lambda_5u^2 - \frac{1}{\sqrt{2}}\frac{v^2}{u}\mu\cbeta\sbeta\,, 
\label{eq:Mrho}
\end{align}
while for $M_\eta$,
\begin{align}
& M^\eta_{11} = -\dfrac{1}{\sqrt{2}}u\mu\dfrac{\sbeta}{\cbeta} \,\quad \quad
 M^\eta_{12} = \dfrac{1}{\sqrt{2}}u\mu \,, \notag \\
& M^\eta_{13} = \dfrac{1}{\sqrt{2}}v\mu\sbeta \,,\quad\quad
 M^\eta_{22} = -\frac{1}{\sqrt{2}}u\mu\dfrac{\cbeta}{\sbeta} \,, \notag \\
& M^\eta_{23}  = -\dfrac{1}{\sqrt{2}}v\mu\cbeta \,,\quad \quad
 M^\eta_{33} = -\frac{1}{\sqrt{2}}\dfrac{v^2}{u}\mu\cbeta\sbeta \,. 
\label{eq:Meta}
\end{align}
All terms in $M^\eta$ are proportional to $\mu$, which explains the need to set  $Y'_{\Phi_1} = Y'_\chi$, otherwise the $\mu$ term in (\ref{eq:V}) would not be present and we would have a massless scalar.

The matrix $M^\eta$ is diagonalised as $R^T M^\eta R = (M^\eta)_\text{diag}$, using a rotation
\begin{equation}
R = \left( \begin{array}{ccc}
-\sbeta & \cbeta & 0 \\ \cbeta & \sbeta & 0 \\ 0 & 0 & 1
\end{array}
\right) 
 \left( \begin{array}{ccc}
-\salpha & 0 & \calpha  \\ 0 & 1 & 0 \\ \calpha & 0 & \salpha
\end{array}
\right) \,,
\end{equation}
being $\alpha$ given by
\begin{equation}
    \tan \alpha = -\frac{u}{v\cbeta\sbeta} \,.
    \label{eq:alpha}
\end{equation}
The only non-zero eigenvalue is
\begin{equation}
m_{A^0}^2 = - \frac{\mu}{\sqrt{2}} \left[\frac{u}{\sbeta \cbeta} + \frac{\sbeta \cbeta v^2}{u} \right]\,,
\label{eq:mA0}
\end{equation}
which corresponds to the squared mass of a CP-odd scalar $A^0$.
The matrix $O$ that rotates the $\rho_i$ fields to the mass basis is parameterised by three mixing angles, $\alpha_1$, $\alpha_2$, $\alpha_3$, and it is given by
\begin{equation}    
O = \begin{pmatrix} c_1c_2 & s_1c_2 & s_2 \\ -c_1s_2s_3 - s_1c_3 & c_1c_3 - s_1s_2s_3 & c_2s_3 \\ -c_1s_2c_3 + s_1s_3 & -c_1s_3 - s_1s_2c_3 & c_2c_3 
\end{pmatrix}^T\,,
\label{eq:O}
\end{equation}
where $c_{1,2,3}=\cos\alpha_{1,2,3}$, $s_{1,2,3}=\sin\alpha_{1,2,3}$ and $-\pi/2 \leq \alpha_{1,2,3} \leq \pi/2$. Let us label as $h$ the SM-like Higgs boson and $H_{1,2}$ the new CP-even scalars, with $H_1$ and $H_2$ being always defined in such a way that $m_{H_1}<m_{H_2}$. We can write $M^\rho$ in terms of the masses of those three scalars and the three mixing angles introduced in Eq.~(\ref{eq:O}). Namely,
\begin{equation}
M^\rho = O\begin{pmatrix} m_h^2 & 0 & 0 \\ 0 & m_{H_1}^2 & 0 \\ 0 & 0 & m_{H_2}^2 \end{pmatrix}O^T\,.
\label{eq:MH}
\end{equation}
Finally, the charged-scalar mass matrix in the basis $(\phi_1^\pm,\phi_2^\pm)$ is
\begin{equation}
    \mathcal{M}^c = \left[v^2\lambda_4 + \frac{\sqrt{2}u\mu}{s_\beta c_\beta}\right]
    \begin{pmatrix} -s_\beta^2 & s_\beta c_\beta \\
    s_\beta c_\beta & -c_\beta^2 \end{pmatrix}\,,
\end{equation}
and its diagonalisation is performed as $U^T \mathcal{M}^c U = (\mathcal{M}^c)_\text{diag}$, with
\begin{equation}
U = \left( \begin{array}{cc}
-\sbeta & \cbeta \\ \cbeta & \sbeta
\end{array}
\right)  \,,
\label{eq:U}
\end{equation}
like in 2HDMs. The non-zero eigenvalue of $\mathcal{M}^c$ is
\begin{equation}
    m_{H^\pm}^2 = -\lambda_4v^2 - \frac{\sqrt{2}u\mu}{\cbeta\sbeta}\,,
    \label{eq:Mch}
\end{equation}
corresponding to the squared mass of new charged scalars $H^\pm$. In Appendix~\ref{sec:A} we show how this result, together with Eqs.~(\ref{eq:mA0}) and (\ref{eq:MH}), can be used to write $\mu$ and $\lambda_{1-7}$ as functions of the physical parameters we have just presented.

Defining $\tilde{H}=(h,H_1,H_2)$, the couplings involving three CP-even scalars can be generically written as
\begin{equation}
\mathcal{L}_{3\tilde{H}} = -v\frac{\lambda_{ijk}}{S_{ijk}}\tilde{H}_i\tilde{H}_j\tilde{H}_k \,,
\label{eq:L3H}
\end{equation}
where the coefficients $\lambda_{ijk}$ are symmetric under index interchange. The symmetry factors $S_{ijk}$ are equal to 1 if all indices are different, 2 if two indices are equal, or 6 if $i=j=k$. The three-scalar interactions involving the pseudoscalar $A^0$ and the charged scalars $H^\pm$ are
\begin{align}
\mathcal{L}_{\tilde{H}A^0A^0} &= -\frac{v}{2}\sum_{i=1}^3 g_{\tilde{H}_iA^0A^0}\tilde{H}_iA^0A^0
\,,\\
\mathcal{L}_{\tilde{H}H^+H^-} &= -v\sum_{i=1}^3 g_{\tilde{H}_iH^+H^-} \tilde{H}_iH^+H^-\,,
\label{eq:LHXX}
\end{align}
respectively. The coefficients $\lambda_{ijk}$, $g_{\tilde{H}_iA^0A^0}$ and $g_{\tilde{H}_iH^+H^-}$ are collected in Appendix~\ref{sec:B}.

The gauge-boson masses and the gauge-scalar interactions can be obtained from the scalar kinetic terms
\begin{equation}
    \mathcal{L} = |D_\mu\Phi_1|^2 + |D_\mu\Phi_2|^2 + |D_\mu\chi|^2\,,
\end{equation}
with the covariant derivatives defined as
\begin{eqnarray}
D_\mu\Phi_1 & = & \left(\partial_\mu - igW_\mu^a\frac{\tau^a}{2} - \frac{1}{2}ig'B_\mu - ig_{Z'}Y'_{\Phi_1} B'_\mu \right) \Phi_1 \,, \notag \\
D_\mu\Phi_2 & = & \left(\partial_\mu - ig W_\mu^a\frac{\tau^a}{2} - \frac{1}{2}ig'B_\mu \right) \Phi_2 \,, \notag \\
D_\mu\chi & = & \left(\partial_\mu - ig_{Z'}Y'_\chi B'_\mu\right) \chi \,.
\end{eqnarray}
As usual, $W_\mu^a$ and $B_\mu$ are the SM gauge fields, while $B'_\mu$ is the one corresponding to the new U(1)$'$ symmetry. Notice that $Y'_{\Phi_1}=Y'_\chi = 9 Y'_q$, as discussed above. As in the SM, the $W$-boson mass at the leading order is
\begin{equation}
m_W^2 = \frac{g^2}{4}v^2\,,
\end{equation}
whereas for the neutral gauge bosons we have, in the $(W_\mu^3\,\,B_\mu\,\,B'_\mu)$ basis, the mass matrix
\begin{equation}
\left( \begin{array}{ccc}
\frac{g^2}{4}v^2 & -\frac{gg'}{4}v^2 & -\frac{gg_{Z'}}{2}Y'_\chi v^2\cbeta^2 \\[4pt] -\frac{gg'}{4}v^2 & \frac{{g'}^2}{4}v^2 & \frac{g'g_{Z'}}{2}Y'_\chi v^2\cbeta^2 \\[4pt] -\frac{gg_{Z'}}{2}Y'_\chi v^2\cbeta^2 & \frac{g'g_{Z'}}{2}Y'_\chi v^2\cbeta^2 & g_{Z'}^2 Y^{\prime 2}_\chi \left(u^2 + v^2\cbeta^2\right) \end{array} 
\right) \,.
\end{equation}
Similarly to the SM, one can write
\begin{equation}
\left( \begin{array}{c} W^3_\mu \\ B_\mu \end{array} \right)
= \left( \begin{array}{cc} s_W & c_W \\ c_W & -s_W \end{array} \right)
\left( \begin{array}{c} A_\mu \\ Z^0_\mu \end{array} \right)
\end{equation}
with $c_W = \cos \theta_W$, $s_W = \sin \theta_W$, $\theta_W$ being the weak mixing angle. The $A_\mu$ field is massless and corresponds to the physical photon. For the two remaining fields the mass matrix is
\begin{equation}
\left( \begin{array}{cc} m_{ZZ}^2 & m_{ZZ'}^2 \\ m_{ZZ'}^2 & m_{Z'Z'}^2 \end{array} \right) \,,
\end{equation}
where
\begin{align}
& m_{ZZ}^2 = \frac{1}{4} g_1^2v^2 = \frac{m_W^2}{\cW^2} \,, \notag \\
& m_{ZZ'}^2 = - \frac{1}{2} g_1 g_{Z'} Y_\chi' v^2 \cbeta^2 \,, \notag \\
& m_{Z'Z'}^2 = (g_{Z'}Y_\chi')^2(u^2+v^2\cbeta^2)\,,
\label{eq:MZZpel}
\end{align}
with $g_1 = g/c_W$. The diagonalisation of this matrix is
\begin{equation}
\left( \begin{array}{c} Z^0_\mu \\ B'_\mu \end{array} \right)
= \left( \begin{array}{cc} c_Z & -s_Z \\ s_Z & c_Z \end{array} \right)
\left( \begin{array}{c} Z_\mu \\ Z^\prime_\mu \end{array} \right) \,,
\end{equation}
with $s_Z = \sin \theta_Z$, $c_Z = \cos \theta_Z$, $\theta_Z$ being the $Z-Z'$ mixing angle.
The tree-level masses of the SM $Z$ boson and the new $Z'$ correspond to the two non-zero eigenvalues,
\begin{eqnarray}
m_{Z,Z'}^2 & = & \frac{1}{2}\left[m_{ZZ}^2 + m_{Z'Z'}^2  \right. \notag \\ 
& & \left. \pm \left((m_{Z'Z'}^2 - m_{ZZ}^2)^2 + 4 m_{ZZ'}^4\right)^{1/2}\right]\,,
\label{eq:MZZpexact}
\end{eqnarray}
 and the mixing angle is given by
\begin{equation}
\tan 2 \theta_Z = \frac{g_1 g_{Z'} Y'_\chi v^2 c_\beta^2}{(g_{Z'} Y'_\chi)^2 (u^2 + v^2 c_\beta^2) - g_1^2 v^2 / 4} \,. 
\end{equation}
In the limit of small $Z-Z'$ mixing, i.e. $\theta_Z\ll 1$, the masses of $Z$ and $Z'$ can be approximated as:
\begin{eqnarray}
m_{Z'}^2 & \simeq & (g_{Z'}Y'_\chi)^2(u^2 + v^2\cbeta^2)\,, \notag \\
m_{Z}^2 & \simeq & \frac{m_W^2}{c_W^2}\left[1 - \frac{(g_{Z'}Y'_\chi)^2 v^2\cbeta^4}{m_{Z'}^2}\right]\,,
\end{eqnarray}
while for $\theta_Z$ one has
\begin{equation}
\theta_Z\simeq \frac{g_{Z'}Y'_\chi m_W v\cbeta^2}{m_{Z'}^2\cW} \,.
\end{equation}

The interaction of the $Z$ boson with fermions receives a small correction due to $Z-Z'$ mixing,
\begin{equation}
\mathcal{L} = g_1 \bar \psi \left[ g_1 (T_3 - s_W^2 Q) c_Z + g_{Z'} Y' s_Z \right] \gamma^\mu \psi Z_\mu \,, 
\end{equation}
with $T_3$ the third isospin component, $Q$ the electric charge and $Y'$ the $\text{U}(1)'$ hypercharge of the field $\psi$. The interaction of the $Z'$ boson with quarks is
\begin{equation}
\mathcal{L} = g_{Z'} \bar q \gamma^\mu (y_L P_L + y_R P_R) q Z'_\mu \,, 
\end{equation}
where the left- and right-handed couplings are
\begin{eqnarray}
y_L & = & Y'_q c_Z - \frac{g_1}{g_{Z'}} (T_3 - Q s_W^2) s_Z \,, \notag \\
y_R & = & Y'_q c_Z + \frac{g_1}{g_{Z'}} Q s_W^2 s_Z \,.
\end{eqnarray}
The lagrangian terms involving two gauge bosons and one scalar can be written as
\begin{eqnarray}
\mathcal{L} & = & g_{ W^+W^- \tilde{H}_i} {W^-}^\mu W^+_\mu \tilde{H}_i + g_{ ZZ \tilde{H}_i} Z^\mu Z_\mu \tilde{H}_i \notag \\
& & + g_{ZZ' \tilde{H}_i}Z^\mu Z'_\mu \tilde{H}_i + (g_{ZW^\pm H^\pm} Z^\mu W^+_\mu H^-
\notag \\
& &  + g_{Z'W^\pm H^\pm} {Z'}^\mu W^+_\mu H^- + {\rm h.c.})
\end{eqnarray}
with
\begin{eqnarray}
g_{W^+W^- \tilde{H}_i} & = & \frac{g^2 v}{2}(O_{1i}\cbeta + O_{2i}\sbeta) \,, \notag \\
g_{ZZ \tilde{H}_i} & = & \frac{g_1^2 v}{4}(O_{1i}\cbeta + O_{2i}\sbeta)\cZ^2 \notag \\
& & - g_1 g_{Z'} Y'_\chi v O_{1i} \cbeta \cZ\sZ \notag \\
& & + g_{Z'}^2 Y^{\prime 2}_\chi (O_{1i}v\cbeta + O_{3i}u)\sZ^2 \,, \notag \\
g_{ZZ' \tilde{H}_i} & = & -\frac{g_1^2 v}{2}(O_{1i}\cbeta + O_{2i}\sbeta)\cZ\sZ \notag \\
& & - g_1 g_{Z'} Y'_\chi v O_{1i}\cbeta(\cZ^2 - \sZ^2) \notag \\
& & +2 g_{Z'}^2 Y^{\prime 2}_\chi (O_{1i}v\cbeta + O_{3i}u)\cZ\sZ \,, \notag \\
g_{ZW^\pm H^\pm} & = & -g g_{Z'} Y'_\chi v \cbeta\sbeta \sZ \,, \notag \\
g_{Z'W^\pm H^\pm} & = & - g g_{Z'} Y'_\chi v \cbeta\sbeta \cZ \,.
\label{eq:SVV}
\end{eqnarray}
The Lagrangian terms with two scalars and one gauge boson are
\begin{eqnarray}
\mathcal{L} & = & g_{Z\tilde{H}_iA^0} \tilde{H}_i\overleftrightarrow{\partial_\mu} A^0 Z^\mu
+ g_{Z'\tilde{H}_iA^0}\tilde{H}_i\overleftrightarrow{\partial_\mu} A^0 {Z'}^\mu
\notag \\
& & + g_{\gamma H^+H^-} {A}^\mu H^+ \overleftrightarrow{\partial_\mu} H^- 
+ g_{ZH^+H^-} {Z}^\mu H^+ \overleftrightarrow{\partial_\mu} H^-
 \notag \\
& & + g_{Z'H^+H^-} {Z'}^\mu H^+ \overleftrightarrow{\partial_\mu} H^- + (g_{ W^\pm \tilde{H}_iH^\pm }{W^+}^\mu H^-\overleftrightarrow{\partial_\mu} \tilde{H}_i \notag \\
& & + g_{W^\pm A^0 H^\pm }{W^+}^\mu H^-\overleftrightarrow{\partial_\mu} A^0 + {\rm h.c.}) \,,
\end{eqnarray}
with
\begin{eqnarray}
g_{Z\tilde{H}_iA^0} & = & \frac{g_1}{2}(O_{1i}\sbeta - O_{2i}\cbeta) \salpha c_Z \notag \\
& & - g_{Z'}Y'_\chi (O_{1i}\sbeta\salpha + O_{3i}\calpha)s_Z \,, \notag \\
g_{Z'\tilde{H}_iA^0} & = & \frac{g_1}{2}(O_{2i}\cbeta - O_{1i}\sbeta)\salpha \sZ \notag \\
& & - g_{Z'} Y'_\chi(O_{1i}\sbeta\salpha + O_{3i}\calpha)\cZ \,, \notag \\
g_{\gamma H^+H^-} & = & i e \,, \notag \\
g_{Z H^+H^-} & = & i \left[g_1 \left(\frac{1}{2}-\sW^2\right)\cZ  + g_{Z'} Y'_\chi \sbeta^2\cZ \right] \,, \notag \\
g_{Z'H^+H^-} & = & i \left[-g_1 \left(\frac{1}{2}-\sW^2\right)\sZ  + g_{Z'}Y'_\chi \sbeta^2\cZ \right] \,, \notag \\
g_{W^\pm \tilde{H}_iH^\pm } & = & i\frac{g}{2} (-O_{1i}\sbeta + O_{2i}\cbeta) \,, \notag \\
g_{W^\pm A^0 H^\pm } & = & \frac{g}{2} s_\alpha \,.
\label{eq:SSV}
\end{eqnarray}
The above interactions allow to compute the different partial decay widths for the $Z'$ boson and the scalars. These are collected in Appendix~\ref{sec:C}.

\section{Parameter space scan}
\label{sec:3}

We use the code \texttt{ScannerS}~\cite{Muhlleitner:2020wwk} to scan the parameter space of the UN2HDM and to check whether the points in parameter space are allowed or excluded at the 95$\%$ confidence level (CL). In this analysis, the following constraints are taken into account:\footnote{From now on we will consider that the new fermions stemming from the vector-like degrees of freedom are heavy enough not to be produced in the decays of the $Z'$ boson. Signals from the new leptons were studied in Ref.~\cite{Aguilar-Saavedra:2019iil}.}
\begin{itemize}
    \item Theoretical constraints imposed by perturbative unitarity, boundedness from below and vacuum stability conditions~\cite{Muhlleitner:2016mzt}. These are applied after calculating the parameters of the scalar potential using the equations collected in Appendix~\ref{sec:A};
    \item Electroweak precision constraints, which use fit results for the oblique parameters $S$, $T$ and $U$ shown in~\cite{Haller:2018nnx}. These are compared with UN2HDM predictions for those parameters, implemented in {\texttt ScannerS} using the results of~\cite{Grimus:2007if,Grimus:2008nb}; 
    \item Flavour constraints based on fit results of~\cite{Haller:2018nnx}, which set limits in the $(m_{H^\pm},\tan\beta)$ plane;
    \item Compatibility of the SM-like scalar with the properties of the experimentally discovered Higgs boson;
    \item Bounds from direct searches for beyond SM scalars.
\end{itemize}
To incorporate the constraints mentioned in the last two points, \texttt{ScannerS} provides an interface to \texttt{HiggsSignals}~\cite{Bechtle:2013xfa,Bechtle:2020uwn} and \texttt{HiggsBounds}~\cite{Bechtle:2008jh,Bechtle:2011sb,Bechtle:2013wla,Bechtle:2015pma,Bechtle:2020pkv}. Some of the inputs required by those two tools are the branching ratios of all scalars. These are computed by the library \texttt{N2HDECAY}~\cite{Engeln:2018mbg}, which is also included in \texttt{ScannerS}\footnote{Due to the differences between the scalar potential of the UN2HDM and the one in~\cite{Engeln:2018mbg}, the triple scalar couplings in Appendix~\ref{sec:B} have to be implemented in \texttt{ScannerS}. In contrast, no changes are done to the couplings between scalars and SM gauge bosons, since they are equal to those found in Eqs.~(\ref{eq:SVV}) and (\ref{eq:SSV}) in the limit $\theta_Z\ll 1$.}.

\begin{table}[htb]
\begin{center}
\begin{tabular}{cc}  \toprule \addlinespace[0.1cm]
 Parameter & Range \\ 
\midrule \addlinespace[0.1cm]
$m_h$ & $125.09~{\rm GeV}$ \\ \addlinespace[0.1cm]
$m_{H_1},m_{H_2},m_{A^0},m_{H^\pm}$ & see Section~\ref{sec:4} \\ 
\midrule \addlinespace[0.1cm]
$\tan\beta$ & $[0,20]$ \\ 
\midrule \addlinespace[0.1cm]
$c(hVV)^2$ & $[0.9,1.0]$ \\ \addlinespace[0.1cm]
$c(ht\overline{t})^2$ & $[0.8,1.2]$ \\ 
\midrule \addlinespace[0.1cm]
${\rm sign}(O_{31})$ & $\{-1,1\}$ \\ \addlinespace[0.1cm]
$O_{32}$ & $[-1,1]$ \\ 
\midrule \addlinespace[0.1cm]
2HDM type & I \\ 
\midrule \addlinespace[0.1cm]
$m_{Z'}$ & $2~{\rm TeV}$ \\ \addlinespace[0.1cm]
$g_{Z'}Y'_q$ & $0.1$ \\ 
\bottomrule
\end{tabular}
\end{center}
\caption{List of common input parameters for the parameter space scan.}
\label{table:input}
\end{table}

In order to increase the efficiency of the scan~\cite{Muhlleitner:2020wwk}, we parameterise the mixing matrix of CP-even scalars using (i) the effective couplings of the SM-like Higgs boson $h$ to top quarks, $c(ht\overline{t})$; (ii) the effective coupling to SM gauge bosons $c(hVV)$, with $V=W,Z$; $O_{32}$ and ${\rm sign}(O_{31})$. Using these four parameters we are able to compute the three mixing angles in $O$, c.f. (\ref{eq:O}) while simultaneously constraining the couplings of $h$ to be SM-like. The type of 2HDM used is also an input, specifying the flavour constraints applied. Finally, the parameters $m_{Z'}$ and $g_{Z'}Y_q'$ are required to extract the VEV of the singlet and the $Z-Z'$ mixing angle. We set a reference mass $M_{Z'} = 2$~TeV. The $Z'$ production cross section is determined by the product $g_{Z'}Y'_q$, which we set to 0.1. The list of common parameters used for all chosen benchmarks is presented in Table~\ref{table:input}, together with their varying ranges. (The ranges for the new scalar masses are different for the various benchmarks examined, as shown in the next section.) With the parameter values shown in Table~\ref{table:input}, we get $u\sim 2.2~{\rm TeV}$ and $\theta_Z<10^{-3}$.\footnote{Notice that this bound on $\theta_Z$ ensures that constraints coming from electroweak precision data are respected since they require typically $\theta_Z \lesssim 10^{-3}$~\cite{Erler:2009jh}.}

For the parameter-space points allowed by the aforementioned constraints, we compute the $Z'$ cross section into SM final states, especially for $Z'\rightarrow W^+W^-$, $Z'\rightarrow Zh$ and $Z'\rightarrow t\overline{t}$, to require agreement with direct searches. This is done using {\scshape MadGraph}~\cite{Alwall:2014hca}, where we consider $Z'$ to be produced from proton-proton collisions with a centre-of-mass energy of 13 TeV. Searches of $Z'$ decaying into dijets are also considered but they are less constraining~\cite{Aguilar-Saavedra:2021qbv}.

\section{Benchmarks}
\label{sec:4}

\begin{figure}
\begin{tabular}{c}
\includegraphics[width=4cm,clip=]{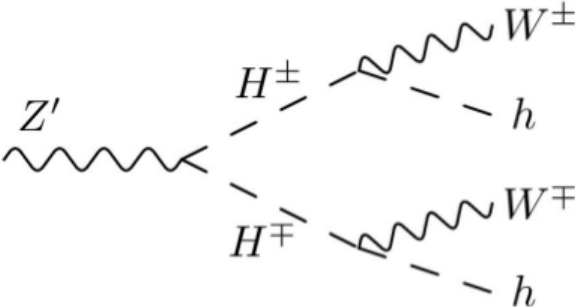} \\[2mm]
\includegraphics[width=4cm,clip=]{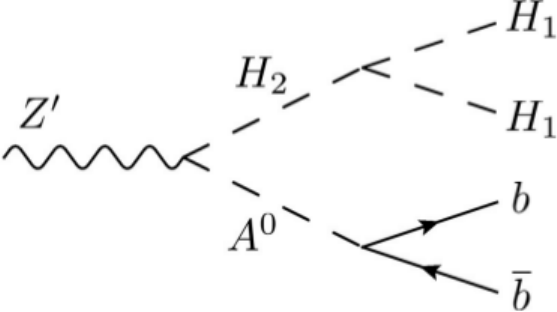} \\[2mm]
\includegraphics[width=4cm,clip=]{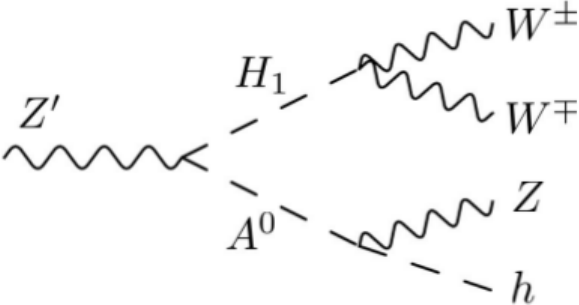}
\end{tabular}
\caption{Feynman diagrams for the $Z'$ cascade decays studied in each benchmark scenario.}
\label{fig:decay}
\end{figure}

Multiboson signals are generated in the UN2HDM by cascade decay of the $Z'$ boson into new scalars, which subsequently decay into $W$, $Z$ bosons or other scalars. We focus on three scenarios that are representative of various types of multiboson signals:
\begin{enumerate} 
\item $Z'\rightarrow H^+H^-$, with $H^\pm\rightarrow W^\pm h$;
\item $Z'\rightarrow H_2A^0$, with $H_2\rightarrow H_1H_1$ and $A^0\rightarrow b\overline{b}$;
\item $Z'\rightarrow H_1A^0$, with $H_1\rightarrow W^+W^-$ and $A^0\rightarrow Zh$;
\end{enumerate}
The Feynman diagrams for these decays are shown in Fig.~\ref{fig:decay}.
In scenario 1, for $m_{H^\pm}$ of few hundreds of GeV, the four bosons in the final state are resolved, yielding a quadriboson signal not experimentally covered. In scenario 2, with $H_2$ and $A_0$ relatively light, the decay $H_2 \to H_1 H_1 \to 4b$ produces a four-pronged jet, while $A_0 \to b \bar b$ produces a two-pronged jet. Finally, scenario 3 with relatively-light $H_1$ and $A^0$ produces two four-pronged jets (when $h^0 \to b \bar b$) with different mass and flavour content. The latter two are partly covered by a search for a heavy resonance decaying into two massive jets~\cite{ATLAS:2020iwa} that unfortunately does not consider jet substructure for the discrimination against SM dijet background.
Table~\ref{table:input2} collects the range for scalar masses used for the scan in each benchmark scenario. 
The scalar branching ratios are computed by \texttt{N2HDECAY}, while for the branching ratios of the $Z'$ boson we use the partial widths collected in Appendix~\ref{sec:C}. In all cases, we consider the new leptons to be heavy enough not to play any role in $Z'$ decays.
\begin{table}[t]
\begin{center}
\begin{tabular}{cccc}
 & Benchmark 1 & Benchmark 2 & Benchmark 3 \\[1mm] 
$m_{H_1}$ & [30,1000] & [20,40] & [150,250] \\[1mm]
$m_{H_2}$ & [30,1000] & [90,110] & [250,1000] \\[1mm]
$m_{A^0}$ & [30,1000] & [90,110] & [200,300] \\[1mm]
$m_{H^\pm}$ & [500,700] & [80,1000] & [80,1000] 
\end{tabular}
\caption{Mass ranges (in GeV) of the new scalars used in each benchmark.}
\label{table:input2}
\end{center}
\end{table}

\subsection{Scenario 1}

\begin{figure}[htb]
\begin{center}
\begin{tabular}{c}
\includegraphics[width=8cm,clip=]{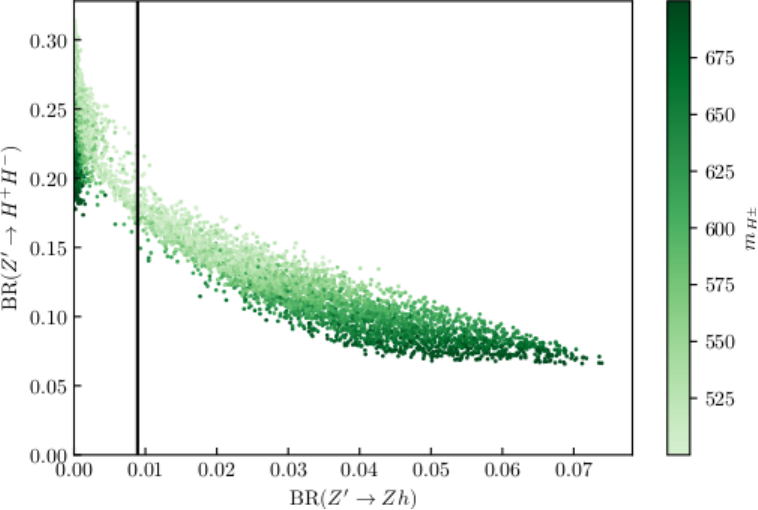} \\[2mm]
\includegraphics[width=8cm,clip=]{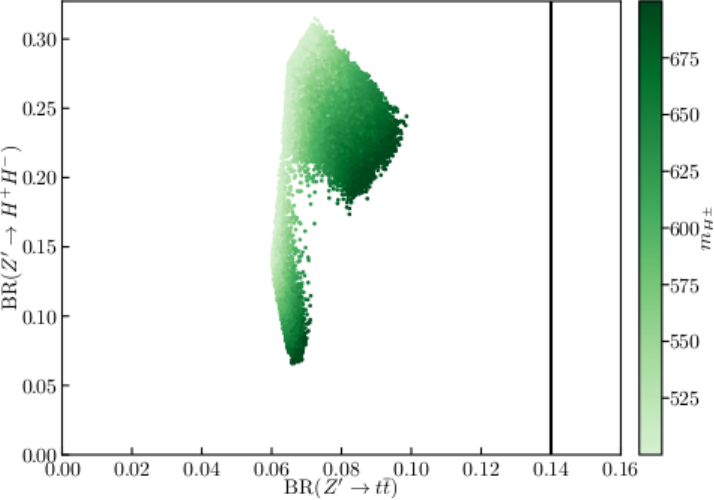}
\end{tabular}
\caption{Branching ratio for $Z' \to H^+ H^-$ versus $Z' \to Z h$ (top) and $Z' \to t \bar t$ (bottom) resulting from the parameter space scan in scenario 1. The vertical lines correspond to the experimental upper bounds. The colour grading is related to the $m_{H^\pm}$ value as shown on the right.}
\label{fig:B1a}
\end{center}
\end{figure}

In this case, the branching ratio for $Z' \to H^+ H^-$ can be up to 0.3, while fulfilling the direct limits on other $Z'$ decay modes (and possible improvements with more data). Fig.~\ref{fig:B1a} (top) shows the branching ratio for $Z' \to H^+ H^-$ versus $Z' \to Z h$. The vertical line corresponds to the experimental upper limit at 95\% CL derived for this mass from current searches~\cite{CMS:2021fyk}, assuming $g_{Z'} Y'_q = 0.1$. A similar plot can be obtained for $Z' \to H^+ H^-$ versus $Z' \to W^+ W^-$, but the allowed area has similar shape and the limit~\cite{ATLAS:2019nat} is less constraining.
It is remarkable that $Z' \to H^+ H^-$ can be sizeable while $Z' \to W^+W^-$ and $Z' \to Zh$ vanish. The reason is that the interactions mediating the latter two modes are proportional to powers of the VEV $v_1$ of the scalar doublet that has non-vanishing hypercharge, either explicitly from $\cos \beta$ factors, or through the $Z-Z'$ mixing. 
The limits from $Z' \to t \bar t$~\cite{ATLAS:2020lks} do not constrain the parameter space allowed by \texttt{ScannerS} (bottom panel), but an improvement by more than a factor of two would exclude the value of $g_{Z'} Y'_q$ used in this benchmark.

\begin{figure}[h]
\begin{center}
\includegraphics[width=8cm,clip=]{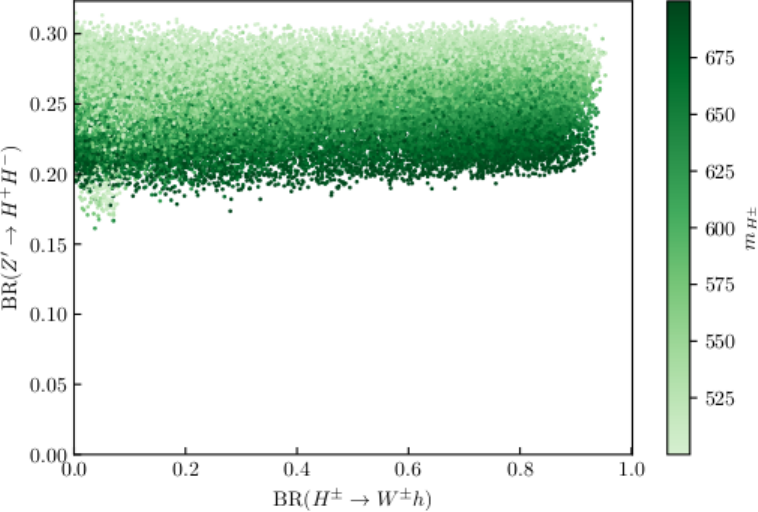}
\caption{Branching ratios for $Z' \to H^+ H^-$ versus $H^\pm \to W^\pm h$ resulting from the parameter space scan  in scenario 1.  The colour grading is related to the $m_{H^\pm}$ value as shown on the right.}
\label{fig:B1b}
\end{center}
\end{figure}
Requiring agreement with direct searches, the branching ratios for $Z' \to H^+ H^-$ versus $H^\pm \to W^\pm h$ are presented in Fig.~\ref{fig:B1b}. One can see that the $H^\pm$ can mostly decay into $W^\pm h$ while having $\text{BR}(Z' \to H^+ H^-) \simeq 0.3$. As a result, the maximum branching ratio for $Z' \to W^+ h W^- h$ reached is 0.27, leading to a cross section times branching ratio of 29 fb for $g_{Z'} Y'_q = 0.1$. 

\subsection{Scenario 2}

\begin{figure}[htb]
\begin{center}
\includegraphics[width=8cm,clip=]{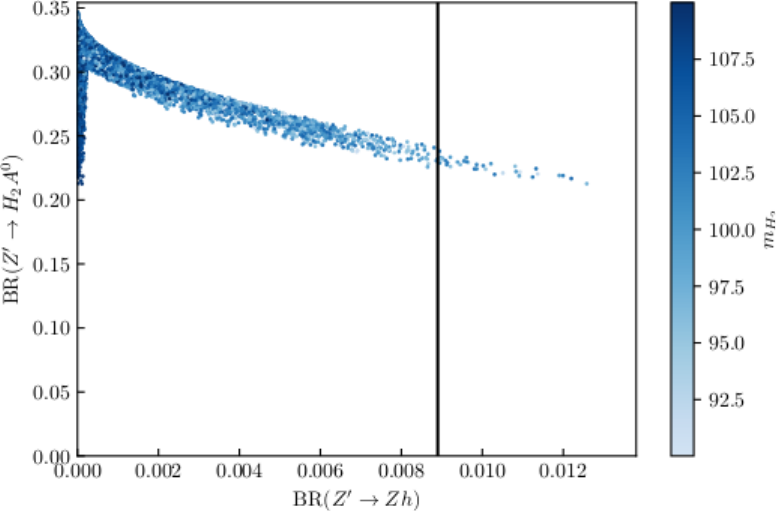} \\[2mm]
\includegraphics[width=8cm,clip=]{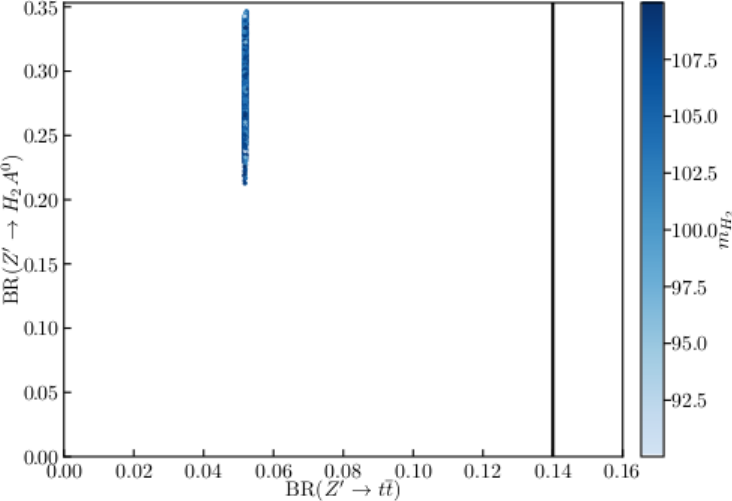}
\caption{Branching ratio for $Z' \to H_2 A^0$ versus $Z' \to Z h$ (top) and $Z' \to t \bar t$ (bottom) resulting from the parameter space scan  in scenario 2. The vertical lines correspond to the experimental upper bound. The colour grading is related to the $m_{H_2}$ value as shown on the right.}
\label{fig:B2a}
\end{center}
\end{figure}

\begin{figure}[htb]
\begin{center}
\includegraphics[width=8cm,clip=]{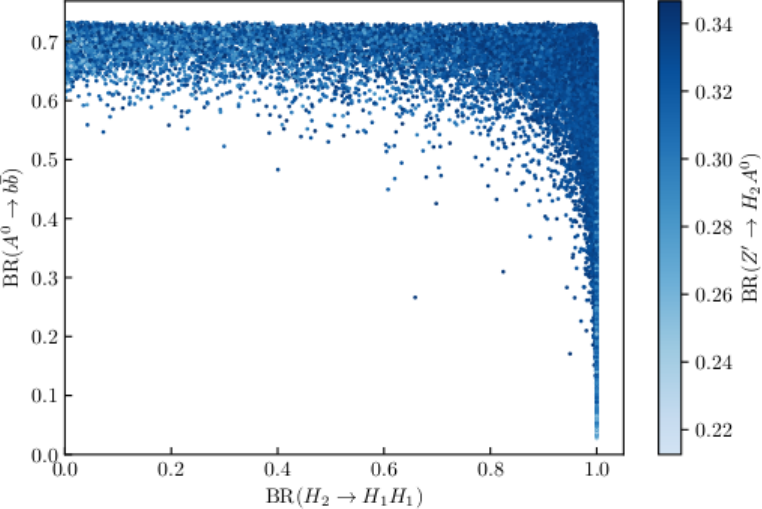}
\caption{Branching ratios for $A^0 \to b \bar b$ versus versus $H_2 \to H_1 H_1$ resulting from the parameter space scan  in scenario 2. The colour grading is related to the value of ${\rm BR}(Z' \to H_2 A^0)$ as shown on the right.}
\label{fig:B2b}
\end{center}
\end{figure}

In this scenario, the branching ratio for $Z' \to H_2 A^0$ can range up to 0.35, while fulfilling the direct limits on other $Z'$ decay modes (and possible improvements with more data). Fig.~\ref{fig:B2a} shows the branching ratio for $Z' \to H_2 A^0$ versus $Z' \to Z h$. The analogous figure considering $Z' \to W^+W^-$ has similar shape, but the current limit does not further constrain the parameter space, and is omitted for brevity. Once more, we observe that the decay into SM bosons $Z' \to W^+W^-$, $Z' \to Zh$ can have negligible rates while $Z' \to H_2 A^0$ is sizeable. This is again because the latter decay is not suppressed when the VEV $v_1$ is small. Direct limits from $Z' \to t \bar t$ do not constrain further the parameter space allowed by \texttt{ScannerS}, as seen in the bottom panel. However, an improvement by more than a factor of two would make this constraint relevant. The results have little dependence on the masses of $H_2$ and $A^0$, which range within a narrow interval $[90,110]$~GeV in this scenario.

The size of the signal in this scenario is determined --- besides the $Z'$ cross section production that is fixed by its mass and $g_{Z'} Y'_q$ --- by the three different branching ratios for $Z' \to H_2 A^0$, $H_2 \to H_1 H_1$ and $A^0 \to b \bar b$. (Since $H_1$ is the lightest scalar, it decays to $b \bar b$ nearly all the time.) We present in Fig.~\ref{fig:B2b} the branching ratios for $A^0 \to b \bar b$ versus $H_2 \to H_1 H_1$, with the colour corresponding to $\text{BR}(Z' \to H_2 A^0)$, for points fulfilling the limit from $Z' \to Zh$. Clearly, all three branching ratios can be sizeable, with a maximum combined branching ratio for $Z' \to H_1 H_1 b \bar b$ of 0.25, leading to a product of the cross section times branching ratio of 27 fb for $g_{Z'} Y'_q = 0.1$. 

\begin{figure}[t!]
\begin{center}
\includegraphics[width=8cm,clip=]{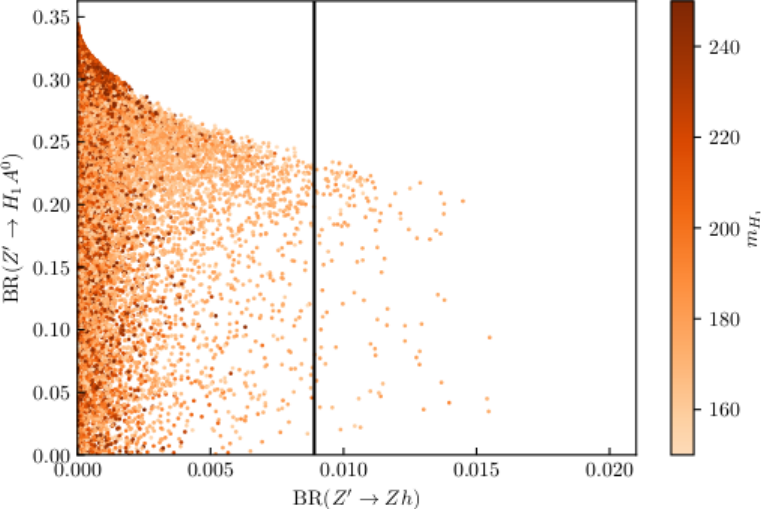} \\[2mm]
\includegraphics[width=8cm,clip=]{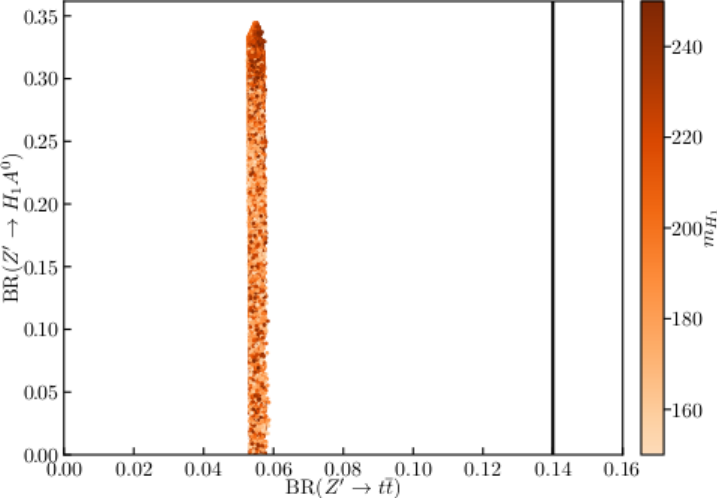} 
\caption{Branching ratio for $Z' \to H_1 A^0$ versus $Z' \to Z h$ (top) and $Z' \to t \bar t$ (bottom) resulting from the parameter space scan  in scenario 3. The vertical lines correspond to the experimental upper bound. The colour grading is related to the $m_{H_2}$ value as shown on the right.}
\label{fig:B3a}
\end{center}
\end{figure}

\subsection{Scenario 3}

The $Z'$ decay mode considered in this scenario is similar to the previous one, but considering instead the lightest new scalar $H_1$, and larger masses for $H_1$ and $A^0$ to allow for other decay modes. The branching ratio for $Z' \to H_1 A^0$ can also range up to 0.35, see Fig.~\ref{fig:B3a} (top). The only direct limit that partially constrains the parameter space allowed by \texttt{ScannerS} is $Z' \to Zh$. As discussed in the previous two benchmarks, future improvements of the limit on $Z' \to t \bar t$ would become constraining (bottom panel).

\begin{figure}[t!]
\begin{center}
\includegraphics[width=8cm,clip=]{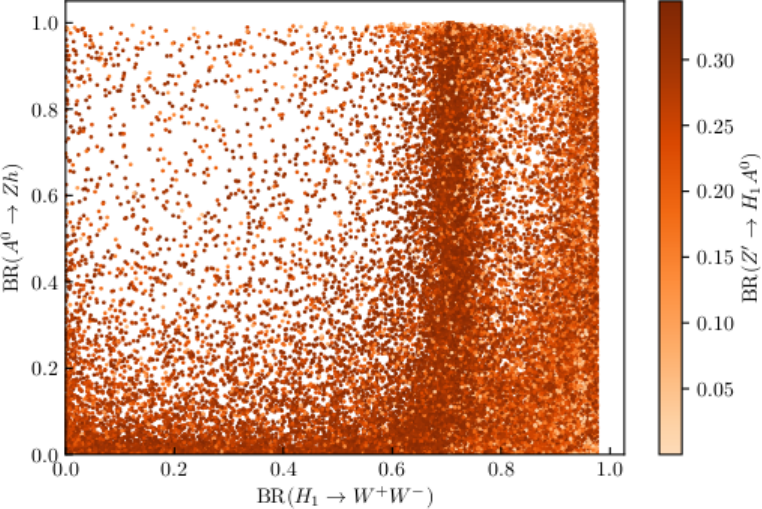}
\caption{Branching ratios for $A^0 \to Zh$ versus versus $H_1 \to WW$ resulting from the parameter space scan  in scenario 3. The colour grading is related to the value of ${\rm BR}(Z' \to H_1 A^0)$ as shown on the right.}
\label{fig:B3b}
\end{center}
\end{figure}

The size of the multiboson signal in this scenario is determined  by the three different branching ratios for $Z' \to H_1 A^0$, $H_1 \to W^+ W^-$ and $A^0 \to Z h$. We present in Fig.~\ref{fig:B3b} the branching ratios for $A^0 \to Z \bar h$ versus $H_1 \to W^+ W^-$, with the colour grading corresponding to the value of $\text{BR}(Z' \to H_1 A^0)$, for points fulfilling the direct limit from $Z' \to Zh$. We point out that the high density of points around $\text{BR}(H_1 \to W^+ W^-) \simeq 0.7$  is motivated by the fact that, when both channels are kinematically open, $\text{BR}(H_1 \to W^+ W^-) \simeq 2 \times \text{BR}(H_1 \to ZZ)$. The latter is suppressed for $H_1$ masses below the $ZZ$ threshold, thus the rate into $W^+ W^-$ can range up to unity for the $H_1$ masses considered. Overall, the three branching ratios can be sizeable, with a maximum combined branching ratio for $Z' \to W^+ W^- Z h$ of 0.29, leading to a cross section times branching ratio of 31 fb for $g_{Z'} Y'_q = 0.1$. In Table~\ref{table:MaxParams} we show the values of scalar masses and $\tan\beta$ of the points with maximum cross section times branching ratio for each scenario.
\begin{table}[t]
\begin{center}
\begin{tabular}{cccc}
 & Benchmark 1 & Benchmark 2 & Benchmark 3 \\[1mm] 
$m_{H_1}$ (GeV) & 223.7 & 35.4 & 178.5 \\[1mm]
$m_{H_2}$ (GeV) & 812.3 & 93.2 & 964.6 \\[1mm]
$m_{A^0}$ (GeV) & 639.1 & 90.0 & 248.0 \\[1mm]
$m_{H^\pm}$ (GeV) & 502.7 & 161.8 & 294.4 \\[1mm]
$\tan\beta$ & 6.6 & 10.3 & 9.4 \\[1mm]
$\sigma\times {\rm BR}$ (fb) & 29 & 27 & 31
\end{tabular}
\caption{Scalar masses and $\tan\beta$ values in the point with highest cross section times branching ratio for each benchmark.}
\label{table:MaxParams}
\end{center}
\end{table}

\section{Discussion}
\label{sec:5}

The LHC has set very stringent limits on the production of new resonances decaying into SM particles, with around 140 fb$^{-1}$ of Run 2 data collected at 13 TeV. Still, many final states remain to be explored. Conspicuously, these unexplored signals might be sizeable, yet compatible with existing limits. We have explored three scenarios for the production of multiboson signals from the decay of a $Z'$ boson. With cross sections times branching ratio around 30 fb in the three scenarios, around 4000 multiboson events could be produced with the collected luminosity.

A wide variety of signal topologies is possible, depending on the decay modes of the $W$, $Z$ and $h$ bosons in the final state. Given the high performance achieved by jet taggers~\cite{Aguilar-Saavedra:2020uhm,Atkinson:2021nlt}, already the hadronic decays of SM bosons (which have the largest branching ratios) are expected to provide a good sensitivity to these types of signals. We note that semileptonic signals are also possible, for example from $H_1 \to W^+ W^-$, when one of the $W$ bosons decays hadronically and the other one leptonically. A detailed study is out of the scope of this work. 

In order to be sensitive to the multiboson signals proposed in this work, dedicated analyses or anomaly-detection methods are required. The signal produced in scenario 2 has a dijet topology, with one jet having four-pronged structure and the other one two-pronged. In scenario 3, when at least one of the $W$ bosons and the $Z/h$ decay hadronically, the signal also has a dijet topology. It has been demonstrated that anomaly detection tools such as CWoLa~\cite{Collins:2018epr,Collins:2019jip} and SOFIE~\cite{Aguilar-Saavedra:2021utu} have the potential to uncover such signals. The signal of scenario 1 has four resolved bosons in the final state. This makes its discrimination from the background technically more demanding, as it features a 4-body resonance plus two intermediate two-body resonances. It is likely that the SOFIE or CATHODE~\cite{Hallin:2021wme} methods are efficient in its detection too.

In summary, the results in this paper show that there are complex new physics signals that could be at reach with already collected LHC data, which motivates the use of generic tools and anomaly-detection strategies to pursue the discovery of any type of physics beyond the SM. We stress that such signals appear naturally in the context of popular SM extensions for which studies are usually focused only on the most simple final-state topologies (mostly direct decays into SM particles).

\section*{Acknowledgements}

We thank Rui Santos and Duarte Azevedo for communications on some aspects related to the \texttt{ScannerS} code. The work of J.A.A.S. has been supported by MICINN project PID2019-110058GB-C21 and CEX2020-001007-S funded by MCIN/AEI/10.13039/501100011033 and by ERDF. F.R.J. and J.F.S. acknowledge Funda\c{c}{\~a}o para a Ci{\^e}ncia e a Tecnologia (FCT, Portugal) for financial support through the projects UIDB/00777/2020, UIDP/00777/2020, CERN/FIS-PAR/0004/2019, and PTDC/FIS-PAR/29436/2017. The work of J.F.S. is supported by the FCT grant SFRH/BD/143891/2019.

\appendix

\section{Higgs potential parameters}
\label{sec:A}

As we have noted in Section~\ref{sec:2}, the scalar potential of the UN2HDM has eleven parameters, which correspond to the same number of physical parameters of the scalar sector of the theory. The VEV of the scalar singlet can be expressed as a function of other physical parameters of the UN2HDM. From (\ref{eq:MZZpexact}) it follows that
\begin{equation}
m_Z^2 + m_{Z'}^2 = m_{ZZ}^2 + m_{Z'Z'}^2\,.
\end{equation}
Using the explicit expressions of these matrix elements in Eq.~(\ref{eq:MZZpel}), we obtain
\begin{equation}
u^2 = \frac{m_Z^2 + m_{Z'}^2 - m_W^2/c_W^2}{(g_{Z'}Y')^2} - v^2\cbeta^2\,.
\end{equation}
By inverting Eq.~(\ref{eq:mA0}), we can write $\mu$ as a function of VEV parameters and the pseudoscalar mass $m_{A^0}$:
\begin{equation}
\mu = -\frac{\sqrt{2}u\cbeta\sbeta}{u^2 + v^2\cbeta^2\sbeta^2}\,m_{A^0}^2\,.
\label{eq:mu}
\end{equation}
The charged-scalar mass in Eq.~(\ref{eq:Mch}) can be used to determine $\lambda_4$,
\begin{equation}
\lambda_4 = -\frac{1}{v^2}\left[m_{H^\pm}^2 + \frac{\sqrt{2}u\mu}{s_\beta c_\beta}\right]\,.
\label{eq:lambda 4}
\end{equation}
The remaining $\lambda_i$ coefficients can be expressed in terms of those parameters using Eq.~(\ref{eq:MH}). Defining $\tilde{M}=(m_h^2,m_{H_1}^2,m_{H_2}^2)$,
\begin{align}
& \lambda_1 = \frac{1}{v^2\cbeta^2}\left[\displaystyle\sum_{i=1}^3 \tilde{M}_iO_{1i}^2 + \frac{u\mu}{\sqrt{2}}\frac{\sbeta}{\cbeta}\right] \,, \notag \\
& \lambda_2 = \frac{1}{v^2\sbeta^2}\left[\displaystyle\sum_{i=1}^3 \tilde{M}_iO_{2i}^2 + \frac{u\mu}{\sqrt{2}}\frac{\cbeta}{\sbeta}\right] \,, \notag \\
& \lambda_3 = \frac{1}{v^2\cbeta\sbeta}\left[\displaystyle\sum_{i=1}^3 \tilde{M}_iO_{1i}O_{2i} - \frac{u\mu}{\sqrt{2}}\right] - \lambda_4 \,, \notag \\
& \lambda_5 = \frac{1}{u^2}\left[\displaystyle\sum_{i=1}^3 \tilde{M}_iO_{3i}^2 + \frac{v^2\mu}{\sqrt{2}u}\cbeta\sbeta\right] \,, \notag \\
& \lambda_6 = \frac{2}{uv\cbeta}\left[\displaystyle\sum_{i=1}^3 \tilde{M}_iO_{1i}O_{3i} - \frac{v\mu}{\sqrt{2}}\sbeta\right] \,, \notag \\
& \lambda_7 = \frac{2}{uv\sbeta}\left[\displaystyle\sum_{i=1}^3 \tilde{M}_iO_{2i}O_{3i} - \frac{v\mu}{\sqrt{2}}\cbeta\right] \,.
\label{eq:lambdas}
\end{align}

\section{Triple scalar couplings}
\label{sec:B}

In the weak basis $w=(\rho_1,\rho_2,\rho_3,\eta_1,\eta_2,\eta_3)$, the lagrangian terms involving three scalar fields can be written as
\begin{equation}
\mathcal{L}_{3H} = - \sum_{p\leq q \leq r}^6 v\,C^{\rm n}_{pqr} w_p w_q w_r\,.
\label{eq:L3Sw}
\end{equation}
The nonzero coefficients $C^{\rm n}_{pqr}$ are given by
\begin{align}
& C^{\rm n}_{111} = C^{\rm n}_{144} = \frac{1}{2}\lambda_1 \cbeta \,, \notag \\
& C^{\rm n}_{112} = C^{\rm n}_{244} = \frac{1}{2}(\lambda_3 + \lambda_4)\sbeta \,, \notag \\
& C^{\rm n}_{113} = C^{\rm n}_{344} = \frac{1}{4}\lambda_6\frac{u}{v} \,, \notag \\
& C^{\rm n}_{122} = C^{\rm n}_{155} = \frac{1}{2}(\lambda_3 + \lambda_4)\cbeta \,, \notag \\
& C^{\rm n}_{123} = C^{\rm n}_{246} = C^{\rm n}_{345} = \frac{1}{v\sqrt{2}}\mu \,, \notag \\
& C^{\rm n}_{133} = C^{\rm n}_{166} = \frac{1}{4}\lambda_6\cbeta \,, \notag \\
& C_{156} = -\frac{1}{v\sqrt{2}}\mu \,, \notag \\
& C^{\rm n}_{222} = C^{\rm n}_{255} = \frac{1}{2}\lambda_2\sbeta \,, \notag \\
& C^{\rm n}_{223} = C^{\rm n}_{355} = \frac{1}{4}\lambda_7\frac{u}{v} \,, \notag \\
& C^{\rm n}_{233} = C^{\rm n}_{266} = \frac{1}{4}\lambda_7\sbeta \,, \notag \\
& C^{\rm n}_{333} = C^{\rm n}_{366} = \frac{1}{2}\lambda_5\frac{u}{v} \,.
\end{align}
The interactions with three scalar fields are thus either of the form
\begin{equation}
\mathcal{L}_{3\tilde{H}} = -\displaystyle\sum_{p\leq q \leq r} v\,C^{\rm n}_{pqr} \rho_p\rho_q\rho_r\,,
\end{equation}
with three CP-even fields, or
\begin{equation}
\mathcal{L}_{\tilde{H}A^0A^0} = -\displaystyle\sum_{p\leq q \leq r} v\,C^{\rm n}_{pqr} \rho_p\eta_{q-3}\eta_{r-3}\,.
\end{equation}
with one CP-even and two CP-odd fields. The weak eigenstates can be written in terms of mass eigenstates as $\rho_i = O_{ia} \tilde{H}_a$ and $\eta_i=R_{i3}A^0$, respectively. Therefore,
\begin{equation}
    \mathcal{L}_{3\tilde{H}} = -\displaystyle\sum v\,C^{\rm n}_{pqr} O_{pa}O_{qb}O_{rc}\tilde{H}_a\tilde{H}_b\tilde{H}_c
\end{equation}
where the sums over $a$, $b$, $c$ and $p\leq q\leq r$ run from 1 to 3, and
\begin{equation}
\mathcal{L}_{\tilde{H}A^0A^0} = -\displaystyle\sum v\,C^{\rm n}_{pqr} O_{pa}R_{q-3,3}R_{r-3,3}\tilde{H}_aA^0A^0\,,
\end{equation}
with $a$ and $p$ running from 1 to 3 and $q\leq r$ from 4 to 6. We write 
\begin{equation}
\lambda_{ijk} = \displaystyle\sum_{p\leq q\leq r,(s)} C^{\rm n}_{pqr}O_{ps_1}O_{qs_2}O_{rs_3}\,,
\label{eq:lijk}
\end{equation}
where $(s)\equiv (s_1,s_2,s_3)$ represents the set of all permutations of the indices $i$, $j$ and $k$. Introducing a symmetry factor $S_{ijk}$ to account for multiple counting of the same terms, we arrive at the expression in (\ref{eq:L3H}). The couplings between a CP-even neutral scalar and two CP-odd ones are given by
\begin{equation}
g_{\tilde{H}_iA^0A^0} = \displaystyle\sum_{p=1}^3\sum_{4\leq q\leq r}^6 v\,C^{\rm n}_{pqr} O_{pi}R_{q-3,3}R_{r-3,3}\,.
\end{equation}
The interactions of one neutral scalar and two charged ones are, in the weak basis,
\begin{equation}
\mathcal{L}_{\tilde{H}H^+H^-} = -\displaystyle\sum_{p=1}^6\sum_{q,r=1}^2 v\,C^{\rm c}_{pqr} w_p\phi^+_q\phi^-_r\,.
\label{eq:LH2Hc}
\end{equation}
The nonzero coefficients $C^{\rm c}_{pqr}$ are
\begin{align}
& C^{\rm c}_{111} = \lambda_1 \cbeta \,, \notag \\
& C^{\rm c}_{112} = C^{\rm c}_{121} = \frac{1}{2}\lambda_4 \sbeta \,, \notag \\
& C^{\rm c}_{122} = \lambda_3 \cbeta \,, \notag \\
& C^{\rm c}_{211} = \lambda_3 \sbeta \,, \notag \\
& C^{\rm c}_{212} = C^{\rm c}_{221} = \frac{1}{2}\lambda_4 \cbeta \,, \notag \\
& C^{\rm c}_{222} = \lambda_2 \sbeta \,, \notag \\
& C^{\rm c}_{311} = \frac{1}{2}\lambda_6 \frac{u}{v} \,, \notag \\
& C^{\rm c}_{312} = C^{\rm c}_{321} = \frac{1}{\sqrt{2}v}\mu \,, \notag \\
& C^{\rm c}_{322} = \frac{1}{2}\lambda_7 \frac{u}{v} \,, \notag \\
& C^{\rm c}_{412} = -C^{\rm c}_{421} = -\frac{i}{2}\lambda_4 \sbeta \,, \notag \\
& C^{\rm c}_{512} = -C^{\rm c}_{521} = \frac{i}{2}\lambda_4 \cbeta \,, \notag \\
& C^{\rm c}_{612} = -C^{\rm c}_{621} = -\frac{i}{\sqrt{2}v}\mu \,.
\end{align}
Using $\phi^\pm_i= U_{i2}H^\pm$, with $U$ in Eq.~(\ref{eq:U}), in the mass-eigenstate basis the terms of Eq.~(\ref{eq:LH2Hc}) that could lead to a coupling between $A^0$, $H^+$ and $H^-$ cancel. On the other hand, the Lagrangian involving one CP-even field and two charged ones can be written as 
\begin{equation}
\mathcal{L}_{\tilde{H}H^+H^-} = -\displaystyle\sum v\,C^{\rm c}_{pqr} O_{pa}U_{q2}U_{r2}\tilde{H}_aH^+H^-\,.
    \label{eq:H2Hc}
\end{equation}
The index $p$ in Eq.~(\ref{eq:H2Hc}) runs from 1 to 3 whereas $q$ and $r$ are either 1 or 2. The couplings of CP-even neutral scalars to charged scalars are given by
\begin{equation}
    g_{\tilde{H}_iH^+H^-} = \displaystyle\sum_{p=1}^3\sum_{q,r=1}^2 v\,C^{\rm c}_{pqr} O_{pi}U_{q2}U_{r2}\,.
    \label{eq: VertexHHchHch}
\end{equation}

\section{Partial widths}
\label{sec:C}

In this appendix we collect for completeness the partial widths for the relevant decay of the new particles introduced in the UN2HDM. The partial widths of the $Z'$ boson are
\begin{align}
& \Gamma(Z'\rightarrow W^+W^-) = \frac{g^2 c_W^2 s_z^2}{192\pi} \frac{m_{Z'}^5}{m_W^4} \left(1-4\frac{m_W^2}{m_{Z'}^2}\right)^{3/2} \notag \\
& \quad \times \left(1 + 20\frac{m_W^2}{m_{Z'}^2} + 12\frac{m_W^4}{m_{Z'}^4}\right)\,, \notag \\
& \Gamma(Z'\rightarrow Z\tilde{H}_i) = \frac{g_{ZZ' \tilde{H}_i}^2}{192\pi m_{Z}^2} \frac{\lambda^{1/2}(m_{Z'}^2,m_Z^2,m_{\tilde{H}_i}^2)}{m_{Z'}} \notag \\
& \quad \times \left[1 + 10 \frac{m_Z^2}{m_{Z'}^2} - 2 \frac{m_{\tilde{H}_i}^2}{m_{Z'}^2} + \frac{m_Z^4}{m_{Z'}^4}
+ \frac{m_{\tilde{H}_i}^4}{m_{Z'}^4} - 2 \frac{m_Z^2 m_{\tilde{H}_i}^2}{m_{Z'}^4}
\right]\,, \notag \\
& \Gamma(Z'\rightarrow W^+ H^-) = \frac{g_{Z'W^\pm H^\pm}^2}{192\pi m_{Z}^2} \frac{\lambda^{1/2}(m_{Z'}^2,m_W^2,m_{H^\pm}^2)}{m_{Z'}} \notag \\
& \quad \left[1 + 10 \frac{m_W^2}{m_{Z'}^2} - 2 \frac{m_{H^\pm}^2}{m_{Z'}^2} + \frac{m_W^4}{m_{Z'}^4}
+ \frac{m_{H^\pm}^4}{m_{Z'}^4} - 2 \frac{m_W^2 m_{H^\pm}^2}{m_{Z'}^4}
\right]\,, \notag \\ 
& \Gamma(Z'\rightarrow \tilde{H}_iA^0) = \frac{g_{Z'\tilde{H}_iA^0}^2}{48\pi m_{Z'}^5} \lambda^{3/2}(m_{Z'}^2, m_{\tilde{H}_i}^2, m_{A^0}^2) \,, \notag \\
& \Gamma(Z'\rightarrow H^+H^-) = \frac{g_{Z'H^+H^-}^2}{48\pi m_{Z'}^5} \lambda^{3/2}(m_{Z'}^2, m_{H^\pm}^2, m_{H^\pm}^2)\,, \notag \\ 
& \Gamma(Z'\rightarrow q\overline{q}) = \frac{N_c g_{Z'}^2 m_{Z'}}{24\pi}
\left( 1-4\frac{m_q^2}{m_{Z'}^2} \right)^{1/2} \notag \\
& \quad \times \left[ (y_L^2 + y_R^2) \left( 1 - \frac{m_q^2}{m_{Z'}^2} \right)
+ 6 y_L y_R \frac{m_q^2}{m_{Z'}^2}
\right] \,,
\end{align}
with the usual kinematical function
\begin{equation}
\lambda(x,y,z) = x^2 + y^2 + z^2 - 2xy - 2xz - 2yz \,.
\end{equation}
The partial widths of the scalars are
\begin{align}
& \Gamma(\tilde{H}_i\rightarrow f \overline{f}) = \frac{N_c}{8\pi} \frac{m_f^2}{v^2 \sin^2\beta} m_{\tilde{H}_i} O_{2i}^2 \left[1 - 4\frac{m_f^2}{m_{\tilde{H}_i}^2}\right]^{3/2}\,, \notag \\
& \Gamma(\tilde{H}_i\rightarrow W^+W^-) = \frac{g_{WW \tilde{H}_i }^2}{64\pi}  \frac{m_{\tilde{H}_i}^3}{m_W^4} \left[1 - 4\frac{m_W^2}{m^2_{\tilde{H}_i}}\right]^{1/2} \notag \\
& \quad \times \left[1 - 4\frac{m_W^2}{m_{\tilde{H}_i}^2} + 12\frac{m_W^4}{m_{\tilde{H}_i}^4}\right] \,, \notag \\
& \Gamma(\tilde{H}_i\rightarrow ZZ) = \frac{g_{ZZ \tilde{H}_i}^2}{32\pi} \frac{m_{\tilde{H}_i}^3}{m_Z^4} \left[1 - 4\frac{m_Z^2}{m^2_{\tilde{H}_i}}\right]^{1/2} \notag \\
& \quad \times \left[1 - 4\frac{m_Z^2}{m_{\tilde{H}_i}^2} + 12\frac{m_Z^4}{m_{\tilde{H}_i}^4}\right] \,, \notag \\
& \Gamma(\tilde{H}_i\rightarrow A^0 Z) = \frac{g_{Z \tilde{H}_iA^0 }^2}{16\pi m_{\tilde{H}_i}^3 m_Z^2} \lambda^{3/2}(m_{\tilde{H}_i}^2,m_{Z}^2,m_{A^0}^2) \,, \notag \\
& \Gamma(\tilde{H}_i\rightarrow W^+ H^-) = \frac{g_{W^\pm \tilde{H}_i H^\pm}^2}{16\pi m_{\tilde{H}_i}^3 m_W^2} \lambda^{3/2}(m_{\tilde{H}_i}^2,m_{W}^2,m_{H^\pm}^2) \,.
\end{align}


\begin{thebibliography}{99}

\bibitem{Aguilar-Saavedra:2015rna}
J.~A.~Aguilar-Saavedra,
JHEP \textbf{10} (2015), 099
[arXiv:1506.06739 [hep-ph]].

\bibitem{Aguilar-Saavedra:2015iew}
J.~A.~Aguilar-Saavedra and F.~R.~Joaquim,
JHEP \textbf{01}, 183 (2016)
[arXiv:1512.00396 [hep-ph]].

\bibitem{ATLAS:2015xom}
G.~Aad et al. [ATLAS Collaboration],
JHEP \textbf{12}, 055 (2015)
[arXiv:1506.00962 [hep-ex]].

\bibitem{CMS:2022lqh}
A.~Tumasyan \textit{et al.} [CMS],
Phys. Rev. Lett. \textbf{129} (2022) no.2, 021802
[arXiv:2201.08476 [hep-ex]].

\bibitem{Agashe:2018leo}
K.~Agashe et al.,
JHEP \textbf{11}, 027 (2018)
[arXiv:1809.07334 [hep-ph]].

\bibitem{Agashe:2017wss}
K.~Agashe et al.,
Phys. Rev. D \textbf{99}, no.7, 075016 (2019)
[arXiv:1711.09920 [hep-ph]].

\bibitem{Aguilar-Saavedra:2017iso}
J.~A.~Aguilar-Saavedra,
JHEP \textbf{05} (2017), 066
[arXiv:1703.06153 [hep-ph]].




\bibitem{ATLAS:2016yqq}
ATLAS Collaboration,
ATLAS-CONF-2016-055.

\bibitem{CMS:2017fgc}
A.~Sirunyan et al. [CMS Collaboration],
Phys. Rev. D \textbf{97}, no.7, 072006 (2018)
[arXiv:1708.05379 [hep-ex]].

\bibitem{Aguilar-Saavedra:2017zuc}
J.~A.~Aguilar-Saavedra,
Eur. Phys. J. C \textbf{77}, no.10, 703 (2017)
[arXiv:1705.07885 [hep-ph]].




\bibitem{Aguilar-Saavedra:2019adu}
J.~A.~Aguilar-Saavedra and F.~R.~Joaquim,
JHEP \textbf{10}, 237 (2019)
[arXiv:1905.12651 [hep-ph]].

\bibitem{Aguilar-Saavedra:2020wrj}
J.~A.~Aguilar-Saavedra and F.~R.~Joaquim,
Eur. Phys. J. C \textbf{80}, no.5, 403 (2020)
[arXiv:2002.07697 [hep-ph]].

\bibitem{CMS:2022shx}
CMS Collaboration,
CMS-PAS-B2G-20-009.




\bibitem{Chen:2013jvg}
C.~Y.~Chen, M.~Freid, M.~Sher,
Phys. Rev. D \textbf{89}, no.7, 075009 (2014)
[arXiv:1312.3949 [hep-ph]].

\bibitem{Muhlleitner:2016mzt}
M.~M\"{u}hlleitner, M.~O.~P.~Sampaio, R.~Santos and J.~Wittbrodt,
JHEP \textbf{03}, 094 (2017)
[arXiv:1612.01309 [hep-ph]].

\bibitem{Keus:2017ioh}
V.~Keus, N.~Koivunen and K.~Tuominen,
JHEP \textbf{09}, 059 (2018)
[arXiv:1712.09613 [hep-ph]].

\bibitem{vonBuddenbrock:2018xar}
S.~von~Buddenbrock et al.,
J. Phys. G \textbf{46}, no.11, 115001 (2019)
[arXiv:1809.06344 [hep-ph]].

\bibitem{Ferreira:2019iqb}
P.~M.~Ferreira, M.~M\"{u}hlleitner, R.~Santos, G.~Weiglein and J.~Wittbrodt,
JHEP \textbf{09}, 006 (2019)
[arXiv:1905.10234 [hep-ph]].

\bibitem{Biekotter:2019kde}
T.~Biek\"{o}tter, M.~Chakraborti and S.~Heinemeyer,
Eur. Phys. J. C \textbf{80}, no.1, 2 (2020)
[arXiv:1903.11661 [hep-ph]].

\bibitem{Engeln:2020fld}
I.~Engeln, P.~M.~Ferreira, M.~M\"{u}hlleitner, R.~Santos, and J.~Wittbrodt,
JHEP \textbf{08}, 085 (2020)
[arXiv:2004.05382 [hep-ph]].

\bibitem{Langacker:2008yv}
P.~Langacker,
Rev. Mod. Phys \textbf{81}, 1199 (2009)
[arXiv:0801.1345 [hep-ph]].

\bibitem{Caron:2018yzp}
S.~Caron, J.~A.~Casas, J.~Quilis and R.~Ruiz de Austri,
JHEP \textbf{12} (2018), 126
[arXiv:1807.07921 [hep-ph]].






\bibitem{Muhlleitner:2020wwk}
M.~M\"uhlleitner, M.~O.~P.~Sampaio, R.~Santos and J.~Wittbrodt,
Eur. Phys. J. C \textbf{82}, no.3, 198 (2022)
[arXiv:2007.02985 [hep-ph]].

\bibitem{Aguilar-Saavedra:2019iil}
J.~A.~Aguilar-Saavedra, J.~A.~Casas, J.~Quilis and R.~Ruiz de Austri,
JHEP \textbf{04} (2020), 069
[arXiv:1911.03486 [hep-ph]].


\bibitem{Haller:2018nnx}
J.~Haller et al.,
Eur. Phys. J. C \textbf{78}, no.8, 675 (2018)
[arXiv:1803.01853 [hep-ph]].

\bibitem{Grimus:2007if}
W.~Grimus, L.~Lavoura, O.~M.~Ogreid and P.~Osland,
J. Phys. G \textbf{35}, 075001 (2008)
[arXiv:0711.4022 [hep-ph]].

\bibitem{Grimus:2008nb}
W.~Grimus, L.~Lavoura, O.~M.~Ogreid and P.~Osland,
Nucl. Phys. B \textbf{801}, 81 (2008)
[arXiv:0802.4353 [hep-ph]].

\bibitem{Bechtle:2013xfa}
P.~Bechtle, S.~Heinemeyer, O.~St\r{a}l, T.~Stefaniak and G.~Weiglein,
Eur. Phys. J. C \textbf{74}, no.2, 2711 (2014)
[arXiv:1305.1933 [hep-ph]].

\bibitem{Bechtle:2020uwn}
P.~Bechtle et al.,
Eur. Phys. J. C \textbf{81}, no.2, 145 (2021)
[arXiv:2012.09197 [hep-ph]].

\bibitem{Bechtle:2008jh}
P.~Bechtle, O.~Brein, S.~Heinemeyer, G.~Weiglein and K.~E.~Williams,
Comput. Phys. Commun. \textbf{181}, 138 (2010)
[arXiv:0811.4169 [hep-ph]].

\bibitem{Bechtle:2011sb}
P.~Bechtle, O.~Brein, S.~Heinemeyer, G.~Weiglein and K.~E.~Williams,
Comput. Phys. Commun. \textbf{182}, 2605 (2011)
[arXiv:1102.1898 [hep-ph]].

\bibitem{Bechtle:2013wla}
P.~Bechtle et al.,
Eur. Phys. J. C \textbf{74}, no.3, 2693 (2014)
[arXiv:1311.0055 [hep-ph]].

\bibitem{Bechtle:2015pma}
P.~Bechtle, S.~Heinemeyer, O.~St\r{a}l, T.~Stefaniak and G.~Weiglein
Eur. Phys. J. C \textbf{75}, no.9, 421 (2015)
[arXiv:1507.06706 [hep-ph]].

\bibitem{Bechtle:2020pkv}
P.~Bechtle et al.,
Eur. Phys. J. C \textbf{80}, no.12, 1211 (2020)
[arXiv:2006.06007 [hep-ph]].

\bibitem{Engeln:2018mbg}
I.~Engeln, M.~M\"uhlleitner and J.~Wittbrodt,
Comput. Phys. Commun. \textbf{234}, 256 (2019)
[arXiv:1805.00966 [hep-ph]]

\bibitem{Erler:2009jh}
J.~Erler, P.~Langacker, S.~Munir and E.~Rojas,
JHEP \textbf{08} (2009), 017
[arXiv:0906.2435 [hep-ph]].

\bibitem{Alwall:2014hca}
J.~Alwall, R.~Frederix, S.~Frixione, V.~Hirschi, F.~Maltoni, O.~Mattelaer, H.~S.~Shao, T.~Stelzer, P.~Torrielli and M.~Zaro,
JHEP \textbf{07}, 079 (2014)
[arXiv:1405.0301 [hep-ph]].

\bibitem{Aguilar-Saavedra:2021qbv}
J.~A.~Aguilar-Saavedra, I.~Lara, D.~E.~Lopez-Fogliani and C.~Mu\~{o}z,
Eur. Phys. J. C \textbf{81}, no.5, 443 (2021)
[arXiv:2101.05565 [hep-ph]]


\bibitem{CMS:2021fyk}
A.~M.~Sirunyan \textit{et al.} [CMS],
Eur. Phys. J. C \textbf{81} (2021) no.8, 688
[arXiv:2102.08198 [hep-ex]].

\bibitem{ATLAS:2019nat}
G.~Aad \textit{et al.} [ATLAS],
JHEP \textbf{09} (2019), 091
[erratum: JHEP \textbf{06} (2020), 042]
[arXiv:1906.08589 [hep-ex]].


\bibitem{ATLAS:2020lks}
G.~Aad \textit{et al.} [ATLAS],
JHEP \textbf{10} (2020), 061
[arXiv:2005.05138 [hep-ex]].

\bibitem{ATLAS:2020iwa}
G.~Aad \textit{et al.} [ATLAS],
Phys. Rev. Lett. \textbf{125} (2020) no.13, 131801
[arXiv:2005.02983 [hep-ex]].

\bibitem{Aguilar-Saavedra:2020uhm}
J.~A.~Aguilar-Saavedra, F.~R.~Joaquim and J.~F.~Seabra,
JHEP \textbf{03} (2021), 012
[arXiv:2008.12792 [hep-ph]].

\bibitem{Atkinson:2021nlt}
O.~Atkinson, A.~Bhardwaj, C.~Englert, V.~S.~Ngairangbam and M.~Spannowsky,
JHEP \textbf{08} (2021), 080
[arXiv:2105.07988 [hep-ph]].

\bibitem{Collins:2018epr}
J.~H.~Collins, K.~Howe and B.~Nachman,
Phys. Rev. Lett. \textbf{121} (2018) no.24, 241803
[arXiv:1805.02664 [hep-ph]].

\bibitem{Collins:2019jip}
J.~H.~Collins, K.~Howe and B.~Nachman,
Phys. Rev. D \textbf{99} (2019) no.1, 014038
[arXiv:1902.02634 [hep-ph]].

\bibitem{Aguilar-Saavedra:2021utu}
J.~A.~Aguilar-Saavedra,
Eur. Phys. J. C \textbf{82} (2022) no.2, 130
[arXiv:2111.02647 [hep-ph]].

\bibitem{Hallin:2021wme}
A.~Hallin, J.~Isaacson, G.~Kasieczka, C.~Krause, B.~Nachman, T.~Quadfasel, M.~Schlaffer, D.~Shih and M.~Sommerhalder,
Phys. Rev. D \textbf{106} (2022) no.5, 055006
[arXiv:2109.00546 [hep-ph]].
\end{thebibliography}
\end{document}